\documentclass[12pt,english]{article}
\usepackage[T1]{fontenc}
\usepackage[latin1]{inputenc}
\usepackage{a4wide}
\usepackage{graphicx}
\usepackage{amssymb}

\makeatletter

\providecommand{\tabularnewline}{\\}

\usepackage{babel}
\makeatother
\begin{document}
\makeatletter

\renewcommand{\theequation}{\hbox{\normalsize\arabic{section}.\arabic{equation}}}

\@addtoreset{equation}{section}

\renewcommand{\thefigure}{\hbox{\normalsize\arabic{section}.\arabic{figure}}}

\@addtoreset{figure}{section}

\renewcommand{\thetable}{\hbox{\normalsize\arabic{section}.\arabic{table}}}

\@addtoreset{table}{section}

\makeatother

\title{{\normalsize\begin{flushright}ITP Budapest Report No. 617\end{flushright}}\vspace{1cm}Finite
size effects in quantum field theories with boundary from scattering
data}

\author{Z. Bajnok$^{1}$%
\footnote{bajnok@afavant.elte.hu%
}, L. Palla$^{2}$%
\footnote{palla@ludens.elte.hu%
}, and G. Takács$^{1}$%
\footnote{takacs@ludens.elte.hu%
}\\
 $^{1}$\emph{\small Theoretical Physics Research Group, Hungarian
Academy of Sciences, }\\
\emph{\small 1117 Budapest, Pázmány Péter sétány 1/A, Hungary}\\
$^{2}$\emph{\small Institute for Theoretical Physics, Eötvös University
}\\
\emph{\small 1117 Budapest, Pázmány Péter sétány 1/A, Hungary}}

\maketitle
\begin{abstract}
We derive a relation between leading finite size corrections for a
1+1 dimensional quantum field theory on a strip and scattering data,
which is very similar in spirit to the approach pioneered by Lüscher
for periodic boundary conditions. The consistency of the results is
tested both analytically and numerically using thermodynamic Bethe
Ansatz, Destri-de Vega nonlinear integral equation and classical field
theory techniques. We present strong evidence that the relation between
the boundary state and the reflection factor one-particle couplings,
noticed earlier by Dorey et al. in the case of the Lee-Yang model
extends to any boundary quantum field theory in 1+1 dimensions.
\end{abstract}

\section{Introduction}

Finite size corrections are important tools in QFT and their use is
mandatory in interpreting numerical data. Nearly two decades ago M.
Lüscher \cite{luscher} gave a general expression for the leading
finite size correction of particle masses and scattering states valid
in any QFT with periodic boundary conditions, in terms of the infinite
volume scattering data. 

Lüscher's result was specialized to 1+1 dimensional field theory \cite{klassen-melzer},
using the exact scattering amplitudes known in integrable field theories.
For example, in a theory with a single particle of mass $m$ and elastic
scattering amplitude $S\left(\theta\right)$ the leading correction
to the particle mass in finite volume is the following:\begin{eqnarray}
\frac{\delta m}{m}\left(L\right) & = & -\frac{\sqrt{3}}{2}i\left(\mathop{\mathrm{Res}}_{\theta=\frac{2\pi i}{3}}S\left(\theta\right)\right)e^{-\frac{\sqrt{3}}{2}mL}\nonumber \\
 & - & \int_{-\infty}^{\infty}\frac{d\theta}{2\pi}\cosh\theta\left(S\left(\theta+\frac{i\pi}{2}\right)-1\right)e^{-mL\cosh\theta}+O\left(e^{-\sqrt{3}mL}\right)\label{eq:original_luscher}\end{eqnarray}
(similar formulae exist for general particle spectra). The first term
(which, if present, dominates over the second) appears whenever the
particle occurs as a bound state of itself, and is proportional to
the residue of the $S$-matrix at the pole, while the second term
is present in general and is of order $L^{-1/2}\mathrm{e}^{-mL}$.

In integrable 1+1 dimensional models restricted to a spacetime strip
of finite width $L$ with integrable boundary conditions, some effort
has been devoted recently to the determination of the leading corrections
to low lying -- mainly ground-state -- energy levels as functions
of $L$ in classical field theory \cite{bpt1,mrs}. However, these
investigations did not look for a general, model independent form
of the asymptotic correction to the infinite volume ground state energy,
which is the goal of the present work to consider. In addition, the
results of the present work will be derived at the quantum level,
and it is shown that they agree to the results in \cite{bpt1,mrs}
in the classical limit.

In this paper we derive a formula, similar in spirit to Lüscher's
result in the case of periodic boundary conditions, which expresses
the finite size correction to the energy of a state containing no
particles (but possibly excited boundary states) in terms of the reflection
amplitudes on the semi-infinite line. Once again, the leading term
depends on the presence of certain poles in the reflection factors.
When such poles are present we get corrections of the form $\mathrm{e}^{-mL}$
(for a particle of mass $m$) with coefficients that depend on the
residue of the respective pole, otherwise the leading correction is
given by a more complicated integral containing the reflection amplitudes
and is of order $L^{-1/2}\mathrm{e}^{-2mL}$. We show that these results
are consistent with the boundary version of thermodynamic Bethe Ansatz
(TBA). Using the general results we determine the form of leading
finite size corrections in boundary sine-Gordon theory, which we check
against results from a nonlinear integral equation (NLIE) in the Dirichlet
case, and compare them to the classical field theory results in \cite{bpt1,mrs}.

\section{Cluster expansion and finite size corrections}

For simplicity let us consider a theory with only one scalar particle
in the bulk. Assume further that the theory is defined on a strip
of width $L$, with boundary conditions labeled by $\alpha$ and $\beta$
such that the elastic reflection amplitudes of the particle on these
boundaries on the semi-infinite line, $R_{\alpha}\left(\theta\right)$
and $R_{\beta}\left(\theta\right)$ are known. To derive the asymptotic
($L\,\rightarrow\,\infty$) form of the ground state energy $E_{\alpha\beta}^{0}(L)$,
we consider the partition function of the theory on a cylinder in
two alternative forms:\[
Z_{\alpha\beta}\left(R,L\right)=\mathrm{Tr}\,\mathrm{e}^{-RH_{\alpha\beta}(L)}= \langle\alpha|\mathrm{e}^{-LH(R)}|\beta\rangle\]
where $H(R)$ denotes the Hamiltonian of the theory in the crossed
channel (on a circle of circumference $R$), $|\alpha\rangle$ and
$|\beta\rangle$ are the appropriate boundary states á la Ghoshal
and Zamolodchikov \cite{gz}, while $H_{\alpha\beta}(L)$ is the Hamiltonian
on the strip. In the limit $R\,\rightarrow\,\infty$ we obtain\[
\mathrm{e}^{-RE_{\alpha\beta}^{0}(L)}\sim\langle\alpha|\mathrm{e}^{-LH(R)}|\beta\rangle\]
The right hand side can be expanded for large $L$ in a so-called
cluster expansion (the analogue of which was recently used in \cite{dfrt}
for the TBA in the other channel to settle long standing issues concerning
the boundary entropy function). Inserting a complete set of eigenstates
of the Hamiltonian $H(R)$ on the circle we obtain\[
\langle\alpha|\mathrm{e}^{-LH(R)}|\beta\rangle=\sum_{n\in\mathcal{H}_{\mathrm{bulk}}}\frac{\langle\alpha|n\rangle\langle n|\beta\rangle}{\langle n|n\rangle}\mathcal{\mathrm{e}}^{-LE_{n}(R)}\]
 For our purposes, only the first few terms are interesting:\begin{eqnarray}
\langle\alpha|\mathrm{e}^{-LH(R)}|\beta\rangle & = & \frac{\langle\alpha|0\rangle\langle0|\beta\rangle}{\langle0|0\rangle}\mathcal{\mathrm{e}}^{-LE_{0}(R)}\nonumber \\
 & + & \sum_{\theta_{n}}\frac{\langle\alpha|\theta_{n}\rangle\langle\theta_{n}|\beta\rangle}{\langle\theta_{n}|\theta_{n}\rangle}\mathcal{\mathrm{e}}^{-L\left(E_{0}(R)+m\cosh\theta_{n}\right)}\nonumber \\
 & + & \sum_{\theta_{n},\theta_{m}}\frac{\langle\alpha|\theta_{n},\theta_{m}\rangle\langle\theta_{n},\theta_{m}|\beta\rangle}{\langle\theta_{n},\theta_{m}|\theta_{n},\theta_{m}\rangle}\mathcal{\mathrm{e}}^{-L\left(E_{0}(R)+m\cosh\theta_{n}+m\cosh\theta_{m}\right)}\nonumber \\
 & + & \mathrm{three\quad particle\quad contributions}+\dots\label{eq:cluster_exp}\end{eqnarray}
In the limit $R\,\rightarrow\,\infty$ the boundary state $|\beta\rangle$
has the general form \cite{gz}\begin{equation}
|\beta\rangle=\left(1+\int_{0}^{\infty}\frac{d\theta}{2\pi}K_{\beta}\left(\theta\right)A^{\dagger}\left(-\theta\right)A^{\dagger}\left(\theta\right)+\dots\right)|0\rangle\label{eq:GZ_bound_state}\end{equation}
(similarly for $|\alpha\rangle$) where the dots denote terms containing
more particles, $A^{\dagger}\left(\theta\right)$ creates an asymptotic
particle of rapidity $\theta$ and \[
K_{\beta}\left(\theta\right)=R_{\beta}\left(\frac{i\pi}{2}-\theta\right)\quad.\]
Normalizing the bulk vacuum energy to $0$ and collecting the leading
(two-particle) terms for large $L$ yields\begin{eqnarray*}
Z_{\alpha\beta}(R,L) & = & 1+mR\int_{0}^{\infty}\frac{d\theta}{2\pi}\cosh\theta\bar{K}_{\alpha}\left(\theta\right)K_{\beta}\left(\theta\right)\mathrm{e}^{-2mL\cosh\theta}+\dots\\
\bar{K}_{\alpha}\left(\theta\right) & = & K_{\alpha}\left(-\theta\right)=R_{\alpha}\left(\frac{i\pi}{2}+\theta\right)\end{eqnarray*}
where we replaced the sums by integrals over the density of states\[
\sum_{\theta_{n}}\rightarrow mR\int\frac{d\theta}{2\pi}\cosh\theta\]
and used real analyticity of the reflection factor\[
R(\theta)^{*}=R(-\theta^{*})\]
For the energy we obtain\begin{equation}
E_{\alpha\beta}(L)=-m\int_{0}^{\infty}\frac{d\theta}{2\pi}\cosh\theta\bar{K}_{\alpha}\left(\theta\right)K_{\beta}\left(\theta\right)\mathrm{e}^{-2mL\cosh\theta}\label{eq:two_part_corr}\end{equation}

One can generalize the above argument to the case when there is a
multiplet of particle of mass $m$. In that case the reflection factor
is a matrix\[
R_{\beta}\left(\theta\right)_{s}^{r}\]
where $r$ and $s$ denote the multiplet labels. The boundary state
takes the form \cite{gz}\[
|\beta\rangle=\left(1+\sum_{r,\, s}\int_{0}^{\infty}\frac{d\theta}{2\pi}K_{\beta}\left(\theta\right)^{rs}A_{r}^{\dagger}\left(-\theta\right)A_{s}^{\dagger}\left(\theta\right)+\dots\right)|0\rangle\]
where\[
K_{\beta}\left(\theta\right)^{r\bar{s}}=R_{\beta}\left(i\frac{\pi}{2}-\theta\right)_{s}^{r}\]
and $\bar{s}$ denotes the charge conjugate of multiplet member $s$.
The leading finite size correction takes the form\begin{equation}
E_{\alpha\beta}(L)=-m\int_{0}^{\infty}\frac{d\theta}{2\pi}\cosh\theta\mathrm{Tr}\,\left(\bar{K}_{\alpha}\left(\theta\right)K_{\beta}\left(\theta\right)\right)\mathrm{e}^{-2mL\cosh\theta}\label{eq:nondiag_two_part_corr}\end{equation}

Let us turn now again to the case of diagonal scattering. If the theory
is integrable then the boundary state is known exactly \begin{equation}
|\beta\rangle=\exp\left(\int_{0}^{\infty}K_{\beta}\left(\theta\right)A^{\dagger}\left(-\theta\right)A^{\dagger}\left(\theta\right)\frac{d\theta}{2\pi}\right)|0\rangle\label{eq:exp1}\end{equation}
and summing up the cluster expansion one can derive a TBA equation
for the ground state energy as done by Leclair et al. in \cite{lmss},
which is analyzed in detail in the next section. The large volume
limit of this TBA is exactly given by the formula (\ref{eq:two_part_corr}). 

However, if there is a pole in the reflection factor at the imaginary
rapidity $\theta=i\pi/2$ of the form\begin{equation}
R_{\beta}\left(\theta\right) \sim \frac{ig_{\beta}^{2}}{2\theta-i\pi}\label{eq:grefl_def}\end{equation}
 the boundary state has an extra one-particle contribution 

\begin{equation}
|\beta\rangle=\left(1+\tilde{g}_{\beta}A^{\dagger}\left(0\right)+\int_{0}^{\infty}\frac{d\theta}{2\pi}K_{\beta}\left(\theta\right)A^{\dagger}\left(-\theta\right)A^{\dagger}\left(\theta\right)+\dots\right)|0\rangle\label{eq:corr_bound_state}\end{equation}
or in the case of an integrable field theory one has \begin{equation}
|\beta\rangle=\left(1+\tilde{g}_{\beta}A^{\dagger}\left(0\right)\right)\exp\left(\int_{0}^{\infty}\frac{d\theta}{2\pi}K_{\beta}\left(\theta\right)A^{\dagger}\left(-\theta\right)A^{\dagger}\left(\theta\right)\right)|0\rangle\label{eq:exp2}\end{equation}
(due to the exclusion principle, adding the one-particle term to the
exponent makes no difference). Ghoshal and Zamolodchikov \cite{gz}
identify $\tilde{g}_{\beta}$ with $g_{\beta}$, but Dorey et al.
in \cite{dptw_onepoint} found the relation\begin{equation}
\tilde{g}_{\beta}=\frac{g_{\beta}}{2}\label{eq:dtw_relation}\end{equation}
comparing one-point functions calculated from the boundary state using
form factor methods to TCSA results in the case of the Lee-Yang model.
As we show below, our results are fully consistent with this latter
suggestion. Namely, using the state (\ref{eq:corr_bound_state}) in
the expansion (\ref{eq:cluster_exp}) one gets for the leading term
the result\[
Z_{\alpha\beta}\left(R,L\right)=1+mR\tilde{g}_{\alpha}\tilde{g}_{\beta}\mathrm{e}^{-mL}+\dots\]
and so\begin{equation}
E_{\alpha\beta}\left(L\right)=-m\frac{g_{\alpha}g_{\beta}}{4}\mathrm{e}^{-mL}+\dots\label{eq:luscher_onepart}\end{equation}
The existence of this contribution requires that both left and right
boundary reflection matrices have a pole at $\theta=i\pi/2$. It is
also important to note that if one-particle terms are present in the
boundary state, the two-particle term of the cluster expansion is
divergent and needs to be regularized. 

Hereafter we shall refer to (\ref{eq:two_part_corr},\ref{eq:luscher_onepart})
as the \emph{boundary Lüscher formulae} because they are a natural
generalization of Lüscher's original results \cite{luscher} to the
case of quantum field theory defined on a strip.

\section{BTBA in the infrared limit}

In order to lend support to relation (\ref{eq:dtw_relation}) between
the one-particle coupling constant appearing in the reflection factor
and the coefficient of the one-particle term in the boundary state,
we calculate the infrared limit of boundary TBA (BTBA). Although the
original derivation given by Leclair et al. \cite{lmss} only considers
the case when there is no one-particle term in the boundary state,
it can be easily argued that the presence of one-particle terms makes
no difference to the end result, and this is also supported by numerical
studies using comparison with truncated conformal space (TCS) by Dorey
et al. in \cite{dptw_btba}. 

For simplicity we consider a theory with a single particle of mass
$m$ on a strip of length $L$. The BTBA equation takes the form\begin{equation}
\epsilon\left(\theta\right)=2mL\cosh\theta-\int_{-\infty}^{\infty}\frac{d\theta'}{2\pi}\Phi\left(\theta-\theta'\right)\log\left(1+\chi\left(\theta'\right)\mathrm{e}^{-\epsilon(\theta')}\right) \label{eq:BTBA_eqn}\end{equation}
where\[
\Phi\left(\theta\right)=-i\frac{d}{d\theta}\log S\left(\theta\right)\]
is the derivative of the phase of the two-particle $S$-matrix $S(\theta)$,
and \[
\chi\left(\theta\right)=\bar{K}_{\alpha}\left(\theta\right)K_{\beta}\left(\theta\right)\]
using the notations of the previous section. Using crossing-unitarity
\cite{gz}\begin{equation}
K_{\alpha}\left(\theta\right)=S\left(2\theta\right)\bar{K}_{\alpha}\left(\theta\right)\label{eq:crossing-unitarity}\end{equation}
and unitarity \[
S(\theta)S(-\theta)=1\]
it is easy to see that $\chi(\theta)$ is an even function.

The energy of the ground state is given by\begin{equation}
E_{\alpha\beta}(L)=-m\int_{-\infty}^{\infty}\frac{d\theta}{4\pi}\cosh\theta\log\left(1+\chi\left(\theta\right)\mathrm{e}^{-\epsilon(\theta)}\right)\label{eq:BTBA_energy}\end{equation}
In the case where there is no one-particle coupling, $\chi(\theta)$
is regular on the real axis and for large $L$ \begin{equation}
\epsilon\left(\theta\right)=2mL\cosh\theta+O\left(\mathrm{e}^{-2mL}\right)\label{eq:epsilon_as}\end{equation}
which gives the following asymptotics for the energy:\[
E_{\alpha\beta}(L)=-m\int_{-\infty}^{\infty}\frac{d\theta}{4\pi}\cosh\left(\theta\right)\chi\theta\mathrm{e}^{-2mL\cosh\theta}+O\left(\mathrm{e}^{-4mL}\right)\]
in perfect agreement with (\ref{eq:two_part_corr}). The same is true
for the case when only one of the boundaries has a nonzero one-particle
coupling. The corresponding $K$ factor has a simple pole at $\theta=0$,
but since the other $K$ is regular and $\chi$ is an even function,
$\chi$ has no singularity at all. This can be understood by recalling
that generally \[
S(0)=-1\]
so crossing-unitarity (\ref{eq:crossing-unitarity}) entails \[
K_{\alpha}(0)=-K_{\alpha}(0)\]
and as a result, $K$ factors must have either a zero or a pole at
$\theta=0$. Even if there is a first order pole in one of the $K$
factors corresponding to a one-particle coupling to the boundary,
the product of the two $K$ factors is still regular.

However, for a theory with nonzero one-particle coupling on both boundaries
$\chi$ has a second-order pole at $\theta=0$. The logarithmic terms
in (\ref{eq:BTBA_eqn}) and (\ref{eq:BTBA_energy}) remain integrable
and the BTBA makes perfectly good sense, but to obtain the correct
asymptotic expansion one needs to be careful. For large $L$, the
energy can be written in the form\begin{equation}
E_{\alpha\beta}(L)=-m\int_{-\infty}^{\infty}\frac{d\theta}{4\pi}\cosh\theta\log\left(1+\chi\left(\theta\right)\mathrm{e}^{-2mL\cosh\theta}\right)\label{eq:asymptotic_integral}\end{equation}
Now we cannot use $\log(1+x)\sim x$ to expand the logarithm as $\chi(\theta)$
diverges at $\theta=0$. In fact, using (\ref{eq:grefl_def}) this
divergence can be computed as \[
\chi\left(\theta\right)\sim\frac{g_{\alpha}^{2}g_{\beta}^{2}}{4\theta^{2}} \sim\frac{g_{\alpha}^{2}g_{\beta}^{2}}{4\sinh^{2}\theta}\]
We can use the following integral formula \cite{gr}\begin{equation}
\int_{-\infty}^{\infty}dx\log\frac{a^{2}+x^{2}}{b^{2}+x^{2}}=2\pi(a-b)\qquad,a,b\geq0\label{eq:integral_formula}\end{equation}
to evaluate the energy in the following way\begin{eqnarray}
E_{\alpha\beta}(L) & = & -m\int_{-\infty}^{\infty}\frac{d\theta}{4\pi}\cosh\theta\left[\log\left(1+\frac{g_{\alpha}^{2}g_{\beta}^{2}}{4\sinh^{2}\theta}\mathrm{e}^{-2mL}\right)+\log\left(\frac{1+\chi\left(\theta\right)\mathrm{e}^{-2mL\cosh\theta}}{1+\frac{g_{\alpha}^{2}g_{\beta}^{2}}{4\sinh^{2}\theta}\mathrm{e}^{-2mL}}\right)\right]\nonumber \\
 & = & -m\frac{\left|g_{\alpha}g_{\beta}\right|}{4}\mathrm{e}^{-mL}-m\int_{-\infty}^{\infty}\frac{d\theta}{4\pi}\cosh\theta\left[\log\left(\frac{1+\chi\left(\theta\right)\mathrm{e}^{-2mL\cosh\theta}}{1+\frac{g_{\alpha}^{2}g_{\beta}^{2}}{4\sinh^{2}\theta}\mathrm{e}^{-2mL}}\right)\right]\label{eq:correct_as}\end{eqnarray}
The remaining integral is an expression of order $\mathrm{e}^{-2mL}$,
while the leading term agrees with (\ref{eq:dtw_relation},\ref{eq:luscher_onepart})
for $g_{\alpha}g_{\beta}>0$. 

It was already noted in \cite{dptw_btba} (using a comparison with
truncated conformal space method) that the BTBA equation only gives
the correct ground state energy in this case, and for values of boundary
parameters such that $g_{\alpha}g_{\beta}<0$ a suitable analytic
continuation must be applied.%
\footnote{Using the notation of paper \cite{dptw_btba}, the regime $g_{\alpha}g_{\beta}>0$
corresponds to the boundary parameter range $-2<b<2$ (for boundary
conditions $\alpha=\mathbb{I}$ and $\beta=\Phi(h)$, with the boundary
parameter $h$ related to $b$), which is the range in which they
found that BTBA compares well with truncated conformal space. For
parameters out of this range the authors needed to resort to analytic
continuation in order to match the numerical data.%
} Note also that in the definition of the boundary coupling $g_{\beta}$
from (\ref{eq:grefl_def}) a branch of the square root function must
be chosen. In all known cases (e.g. Lee-Yang in \cite{dptw_btba}
or sine-Gordon in the following section) there exists a branch choice
such that the boundary couplings depend analytically on the boundary
parameters. In this case a straightforward analytic continuation from
the range of parameters for which $g_{\alpha}g_{\beta}>0$ gives the
result \[
E_{\alpha\beta}(L)=-m\frac{g_{\alpha}g_{\beta}}{4}\mathrm{e}^{-mL}+\dots\]
 in the parameter range where $g_{\alpha}g_{\beta}<0$, which is consistent
with (\ref{eq:dtw_relation},\ref{eq:luscher_onepart}). Furthermore,
the result (\ref{eq:correct_as}) also yields a formula for the regularized
two-particle contribution. The two-particle contribution is regularized
by the logarithm in (\ref{eq:asymptotic_integral}), which represents
a partial re-summation of the higher multi-particle contributions,
and the bulk interaction plays no role at leading order in large volume.
We conjecture that the leading asymptotics calculated from (\ref{eq:correct_as})
is valid also for non-integrable theories, since for large $L$ the
leading contribution comes from a very small region around $\theta=0$
and for such small values of the rapidity the inelastic scattering
channels should play no role provided $m$ is the smallest particle
mass in the spectrum.

\section{Application to sine-Gordon theory with integrable boundary conditions}

Next we apply these results to sine-Gordon theory on a strip with
integrable boundary conditions of Ghoshal-Zamolodchikov type \cite{gz}.
The model has a complicated spectrum with non-diagonal bulk and boundary
scattering. However, to derive the leading term (\ref{eq:two_part_corr},\ref{eq:luscher_onepart})
we used only (\ref{eq:GZ_bound_state},\ref{eq:corr_bound_state})
which is valid in general field theories (even non-integrable ones)
in two dimensions, as follows from the derivation of \cite{gz}. Integrability
is only needed to derive the exact exponentiated form of the boundary
state (\ref{eq:exp1},\ref{eq:exp2}). Besides that, the derivation
of Section 2 carries over to the case of non-diagonal boundary scattering
in a straightforward manner, while the effects of bulk scattering
appear only in higher correction terms.

The action of the model is\begin{equation}
\mathcal{A}=\int_{-\infty}^{\infty}dt\left[\int_{0}^{L}dx\:\mathcal{L}_{SG}(x,t)-V_{B}^{(0)}\left(\Phi(0,t)\right)-V_{B}^{(L)}\left(\Phi(L,t)\right)-\alpha\frac{\partial\Phi(L,t)}{\partial t}\right]\label{eq:action}\end{equation}
where the bulk Lagrangian density is given by\[
\mathcal{L}_{SG}(x,t)=\frac{1}{2}\left(\partial_{\mu}\Phi(x,t)\right)^{2}-\frac{m^{2}}{\beta^{2}}\left(1-\cos\beta\Phi\right)\]
and the boundary interaction is described by the potentials\begin{equation}
V_{B}^{(z)}\left(\Phi(z,t)\right)=M_{z}\left[1-\cos\frac{\beta}{2}\left(\Phi(z,t)-\Phi_{z}\right)\right]\qquad,\qquad z=0,\, L\label{VB}\end{equation}
 (the $M_{z}\,\rightarrow\,\infty$ limit corresponds to Dirichlet
boundary conditions). The last term (proportional to $\alpha$) is
allowed by integrability, and on the strip -- in contrast to the theory
on the half line -- it cannot be eliminated from the action by an
appropriate {}``gauge'' transformation \cite{an}. The spectrum
is constituted by the solitons and breathers, and the scattering theory
on the half line is known. The reflection amplitudes on the two boundaries
can be parameterized with the Ghoshal-Zamolodchikov parameters $\eta_{z}$
and $\vartheta_{z}$ ($z=0,\: L$). 

Using a perturbed conformal field theory description one can write
the action of the quantum theory as \begin{eqnarray*}
\mathcal{A}_{\mathrm{pCFT}} & = & \mathcal{A}_{\mathrm{c=1}}+\mu\int_{-\infty}^{\infty}dt\int_{-\infty}^{0}dx:\cos\beta\Phi:+\tilde{\mu}_{0}\int_{-\infty}^{\infty}dt:\cos\frac{\beta}{2}\left(\Phi(0,t)-\Phi_{0}\right):\\
 & + & \tilde{\mu}_{L}\int_{-\infty}^{\infty}dt:\cos\frac{\beta}{2}\left(\Phi(L,t)-\Phi_{L}\right):\\
\mathcal{A}_{\mathrm{c=1}} & = & \int_{-\infty}^{\infty}dt\int_{-\infty}^{0}dx\frac{1}{2}\left(\partial_{\mu}\Phi(x,t)\right)^{2}-\alpha\int_{-\infty}^{\infty}dt\frac{\partial\Phi(L,t)}{\partial t}\end{eqnarray*}
The parameters of this action are related to the boundary parameters
$\eta$, $\vartheta$ by the UV-IR relation \cite{uv-ir}\begin{eqnarray}
\frac{\tilde{\mu}_{z}}{\tilde{\mu}_{\mathrm{crit}}}\sin\frac{\beta\Phi_{z}}{2} & = & -(-1)^{z/L}\sin\frac{\eta_{z}}{\lambda+1}\sinh\frac{\vartheta_{z}}{\lambda+1}\nonumber \\
\frac{\tilde{\mu_{z}}}{\tilde{\mu}_{\mathrm{crit}}}\cos\frac{\beta\Phi_{z}}{2} & = & \cos\frac{\eta_{z}}{\lambda+1}\cosh\frac{\vartheta_{z}}{\lambda+1}\label{eq:UVIR_relation}\end{eqnarray}
where\begin{equation}
\tilde{\mu}_{\mathrm{crit}}=\sqrt{\frac{2\mu}{\sin\frac{\beta^{2}}{8}}}\quad,\quad\lambda=\frac{8\pi}{\beta^{2}}-1\label{eq:lamdadef}\end{equation}
and the sign factor $-(-1)^{z/L}$ arises from the fact that on the
left end ($z=0$) of the strip the role of the soliton and antisoliton
are interchanged with respect to the right end ($z=L$). In the classical
limit $\tilde{\mu}_{z}$ can be replaced by $M_{z}$ and $\mu$ by
$m^{2}/\beta^{2}$ from (\ref{eq:action}), and the $\eta,\,\vartheta$
parameters must be scaled as \cite{kp}\begin{equation}
\eta=\eta_{\mathrm{cl}}\left(1+\lambda\right)\quad,\quad\vartheta=\vartheta_{\mathrm{cl}}\left(1+\lambda\right)\label{eq:eta_theta_scaling}\end{equation}
keeping $\eta_{\mathrm{cl}},\,\vartheta_{\mathrm{cl}}$ finite. This
leaves us with the classical UV-IR relations\begin{eqnarray}
\frac{M_{z}}{M_{\mathrm{crit}}}\sin\frac{\beta\Phi_{z}}{2} & =- & (-1)^{z/L}\sin\eta_{\mathrm{cl}}\sinh\vartheta_{\mathrm{cl}}\nonumber \\
\frac{M_{z}}{M_{\mathrm{crit}}}\cos\frac{\beta\Phi_{z}}{2} & = & \cos\eta_{\mathrm{cl}}\cosh\vartheta_{\mathrm{cl}}\label{eq:classical_uvir}\\
M_{\mathrm{crit}} & = & \frac{4m}{\beta^{2}}\nonumber \end{eqnarray}

In the regime $\beta^{2}<4\pi$ the spectrum contains the first breather,
which has the reflection matrix \cite{ghoshal}\begin{equation}
R_{|\rangle}^{(1)}(\theta)_{z}=\frac{\left(\frac{1}{2}\right)\left(\frac{1}{2\lambda}+1\right)}{\left(\frac{1}{2\lambda}+\frac{3}{2}\right)}\frac{\left(\frac{\eta_{z}}{\pi\lambda}-\frac{1}{2}\right)\left(\frac{i\vartheta_{z}}{\pi\lambda}-\frac{1}{2}\right)}{\left(\frac{\eta_{z}}{\pi\lambda}+\frac{1}{2}\right)\left(\frac{i\vartheta_{z}}{\pi\lambda}+\frac{1}{2}\right)}\quad,\quad(x)=\frac{\sinh\left(\frac{\theta}{2}+i\frac{\pi x}{2}\right)}{\sinh\left(\frac{\theta}{2}-i\frac{\pi x}{2}\right)}\,.\label{b1refl}\end{equation}
 The breather reflection amplitudes are all independent of the parameter
$\alpha$, which appears only in the soliton reflection matrices.
The reflection factor (\ref{b1refl}) does have a pole at $\theta=i\pi/2$
with the singular term taking the form\[
R_{|\rangle}^{(1)}(\theta)\sim4i\frac{1+\cos\frac{\pi}{2\lambda}-\sin\frac{\pi}{2\lambda}}{1-\cos\frac{\pi}{2\lambda}+\sin\frac{\pi}{2\lambda}}\tan^{2}\frac{\eta}{2\lambda}\tanh^{2}\frac{\vartheta}{2\lambda}\frac{1}{2\theta-i\pi}\]
which gives the following result for the coupling of the first breather
to the boundary with parameters $\eta$ and $\vartheta$\begin{equation}
g_{1}\left(\eta,\vartheta\right)=2\sqrt{\frac{1+\cos\frac{\pi}{2\lambda}-\sin\frac{\pi}{2\lambda}}{1-\cos\frac{\pi}{2\lambda}+\sin\frac{\pi}{2\lambda}}}\tan\frac{\eta}{2\lambda}\tanh\frac{\vartheta}{2\lambda}\label{eq:b1_coupling}\end{equation}
(The expression under the square root is always positive as long as
$\lambda>1$ which is necessary for the first breather to exist in
the spectrum). However, some care must be taken, because the sign
of the coupling $g_{1}$ must be opposite on the two ends of the strip
since these are related by a spatial reflection under which the first
breather is odd. 

As a result, formula (\ref{eq:luscher_onepart}) together with (\ref{eq:dtw_relation})
predicts that in the regime $\beta^{2}<4\pi$ ($\lambda>1$) for generic
boundary conditions the leading finite size correction to the ground
state on the strip is given by\begin{eqnarray}
E_{\alpha\beta}\left(L\right) & = & -\frac{1}{4}m_{1}g_{1}\left(\eta_{L},\vartheta_{L}\right)\left(-g_{1}\left(\eta_{0},\vartheta_{0}\right)\right)\mathrm{e}^{-m_{1}L}\label{eq:m1_leading}\\
 & = & m_{1}\frac{1+\cos\frac{\pi}{2\lambda}-\sin\frac{\pi}{2\lambda}}{1-\cos\frac{\pi}{2\lambda}+\sin\frac{\pi}{2\lambda}}\tan\frac{\eta_{0}}{2\lambda}\tanh\frac{\vartheta_{0}}{2\lambda}\tan\frac{\eta_{L}}{2\lambda}\tanh\frac{\vartheta_{L}}{2\lambda}\mathrm{e}^{-m_{1}L}+\dots\nonumber \end{eqnarray}
where $m_{1}$ is the mass of the first breather, $\eta_{0},\vartheta_{0}$
and $\eta_{L},\vartheta_{L}$ are the boundary parameters characterizing
the boundary at the left and right ends of the strip, respectively.
Soliton corrections are always subleading, since there is no one-particle
coupling of solitons to the boundary and $m_{1}<2M$.

It is also interesting to compute the one-particle term contributed
by the second breather, since in the regime where the second breather
exists ($\beta^{2}<8\pi/3$ i.e. $\lambda>2$), its mass $m_{2}=2m_{1}\cos\frac{\pi}{2\lambda}$
is lower than $2m_{1}$ and so it dominates over the two-particle
correction from the first breather, giving the next-to-leading finite
size correction. Furthermore, if any of the $\eta_{z}\:,\vartheta_{z}$
($z=0,\: L$) parameters is zero, then the leading term (\ref{eq:m1_leading})
vanishes, leaving the second breather's one-particle term as the leading
one.

The reflection amplitude for the second breather is of the form \cite{ghoshal}\begin{eqnarray}
R_{|\rangle}^{(2)}(\theta) & = & \frac{\left(\frac{1}{2}\right)\left(\frac{1}{2\lambda}+1\right)\left(\frac{1}{\lambda}+1\right)\left(\frac{1}{2\lambda}\right)}{\left(\frac{1}{2\lambda}+\frac{3}{2}\right)^{2}\left(\frac{1}{\lambda}+\frac{3}{2}\right)}\nonumber \\
 & \times & \frac{\left(\frac{\eta_{z}}{\pi\lambda}-\frac{1}{2}-\frac{1}{2\lambda}\right)\left(\frac{i\vartheta_{z}}{\pi\lambda}-\frac{1}{2}-\frac{1}{2\lambda}\right)}{\left(\frac{\eta_{z}}{\pi\lambda}+\frac{1}{2}-\frac{1}{2\lambda}\right)\left(\frac{i\vartheta_{z}}{\pi\lambda}+\frac{1}{2}-\frac{1}{2\lambda}\right)}\frac{\left(\frac{\eta_{z}}{\pi\lambda}-\frac{1}{2}+\frac{1}{2\lambda}\right)\left(\frac{i\vartheta_{z}}{\pi\lambda}-\frac{1}{2}+\frac{1}{2\lambda}\right)}{\left(\frac{\eta_{z}}{\pi\lambda}+\frac{1}{2}+\frac{1}{2\lambda}\right)\left(\frac{i\vartheta_{z}}{\pi\lambda}+\frac{1}{2}+\frac{1}{2\lambda}\right)}\label{eq:b2refl}\end{eqnarray}
from which the boundary coupling can be extracted\begin{equation}
g_{2}\left(\eta,\vartheta\right)=\frac{2\tan\left(\frac{\pi}{4\lambda}-\frac{\eta}{2\lambda}\right)\tan\left(\frac{\pi}{4\lambda}+\frac{\eta}{2\lambda}\right)\tan\left(\frac{\pi}{4\lambda}-\frac{i\vartheta}{2\lambda}\right)\tan\left(\frac{\pi}{4\lambda}+\frac{i\vartheta}{2\lambda}\right)}{\tan\frac{\pi}{4\lambda}\sqrt{\tan\frac{\pi}{2\lambda}\tan\left(\frac{\pi}{4}+\frac{\pi}{2\lambda}\right)}}\label{eq:b2_coupling}\end{equation}
and the one-particle contribution to the finite size corrections becomes\begin{equation}
-m_{2}\frac{g_{2}\left(\eta_{0},\vartheta_{0}\right)}{2}\frac{g_{2}\left(\eta_{L},\vartheta_{L}\right)}{2}\mathrm{e}^{-m_{2}L}\label{eq:m2_leading}\end{equation}
(here the sign of the couplings are the same at the two boundaries
as the second breather is even). We remark that in this model there
is in fact no sign ambiguity in this coupling once the signs of bulk
three-particle couplings are fixed, because the soliton reflection
factors contain a pole at  $\theta=i\frac{n\pi}{2\lambda}$, $n=1,2,\dots,[\lambda]$
whose residue is proportional to \[
f_{+-}^{n}g_{n}\left(\eta,\vartheta\right)\]
where $f_{+-}^{n}$ is the bulk soliton-antisoliton-breather coupling,
while $g_{n}\left(\eta,\vartheta\right)$ is the one-particle coupling
of the $n$th breather to the boundary. As $g_{n}\left(\eta,\vartheta\right)$
appears linearly, it can be unambiguously calculated and the result
matches the formulae (\ref{eq:b1_coupling},\ref{eq:b2_coupling}).
We note that in the limit $\vartheta\rightarrow\infty$ (Dirichlet
boundary conditions) the couplings simplify to\begin{eqnarray}
g_{1}\left(\eta,\infty\right) & = & 2\sqrt{\frac{1+\cos\frac{\pi}{2\lambda}-\sin\frac{\pi}{2\lambda}}{1-\cos\frac{\pi}{2\lambda}+\sin\frac{\pi}{2\lambda}}}\tan\frac{\eta}{2\lambda}\nonumber \\
g_{2}\left(\eta,\infty\right) & = & \frac{2\tan\left(\frac{\pi}{4\lambda}-\frac{\eta}{2\lambda}\right)\tan\left(\frac{\pi}{4\lambda}+\frac{\eta}{2\lambda}\right)}{\tan\frac{\pi}{4\lambda}\sqrt{\tan\frac{\pi}{2\lambda}\tan\left(\frac{\pi}{4}+\frac{\pi}{2\lambda}\right)}}\label{eq:dirichlet_couplings}\end{eqnarray}

\section{Comparison with the Dirichlet NLIE }

In this section we check the boundary L\"{u}scher formulae derived
previously by comparing them to the exact ground state energies at
large but finite $L$-s in the sine-Gordon model with Dirichlet boundary
conditions. We determine the exact ground state energies from the
(ground state version of the) NLIE proposed recently in \cite{abr},
which generalizes \cite{lmss} by allowing the sine-Gordon field to
take two different values at the two boundaries of the strip.

The NLIE is a complex nonlinear integral equation of the form\begin{eqnarray}
Z\left(\theta\right) & = & 2ML\sinh\theta+P\left(\theta\,|\, H_{0},H_{L}\right)-i\int_{-\infty}^{\infty}dxG(\theta-x-i\eta)\log\left(1-\mathrm{e}^{iZ(x+i\eta)}\right)\nonumber \\
 &  & +i\int_{-\infty}^{\infty}dxG(\theta-x+i\eta)\log\left(1-\mathrm{e}^{iZ(x-i\eta)}\right)\label{eq:DNLIE}\end{eqnarray}
where\begin{eqnarray*}
P\left(\theta\,|\, H_{0},H_{L}\right) & = & 2\pi\int_{0}^{\theta}dx\left(F\left(x,H_{0}\right)+F\left(x,H_{L}\right)+G(x)+J(x)\right)\\
G(\theta) & = & \int_{-\infty}^{\infty}\frac{dk}{2\pi}\mathrm{e}^{ik\theta}\frac{  \sinh\left(\frac{\pi(1-\lambda)}{2\lambda}k\right)}{\sinh\frac{\pi}{2\lambda}k\cosh\frac{\pi}{2}k}\\
J(\theta) & = & \int_{-\infty}^{\infty}\frac{dk}{2\pi}\mathrm{e}^{ik\theta}\frac{  \sinh\left(\frac{\pi(1-\lambda)}{4\lambda}k\right)\cosh\left(\frac{\pi(1+\lambda)}{4\lambda}k\right)}{\sinh\frac{\pi}{2\lambda}k\cosh\frac{\pi}{2}k}\\
F(\theta,H) & = & \int_{-\infty}^{\infty}\frac{dk}{2\pi}\mathrm{e}^{ik\theta}\frac{\sinh(\frac{\pi}{2}\left(1+\frac{1}{\lambda}-H\right)k)  }{2\sinh\frac{\pi}{2\lambda}k\cosh\frac{\pi}{2}k}\end{eqnarray*}
$M$ is the soliton mass, and $\lambda$ is the parameter defined
in (\ref{eq:lamdadef}). The parameters $H_{0,L}$ can be expressed
from the boundary values of the sine-Gordon field $\Phi$:\[
H_{0}=\frac{1-\frac{8}{\beta}\Phi_{0}}{\lambda}\qquad,\qquad H_{L}=\frac{1+\frac{8}{\beta}\Phi_{L}}{\lambda}\]
and the equation in the present form is valid only for $0<H_{0,L}<1+\lambda^{-1}$.

This equation must be solved for the unknown function $Z(\theta)$.
All the functions in the equations are analytic in some strip containing
the real $\theta$ axis, and $\eta$ must be chosen to lie inside
that strip. Once $Z(\theta)$ is obtained (usually by a straightforward
iteration procedure), the ground state energy of the theory can be
calculated using\begin{equation}
E(L)=-M\:\Im\mathrm{m}\int_{-\infty}^{\infty}\frac{dx}{2\pi}\sinh(x+i\eta)\log\left(1-\mathrm{e}^{iZ(x+i\eta)}\right)\label{eq:NLIE_energy}\end{equation}
It satisfies\[
E(L)\rightarrow0\qquad\mathrm{as}\qquad L\rightarrow\infty\]
which means that the vacuum energy extracted from the NLIE is normalized
to zero vacuum energy density and zero boundary energy in infinite
volume. The true vacuum energy (calculated e.g. in the perturbed conformal
field theory framework) has the form\[
E_{\mathrm{PCT}}=\mathcal{E}_{\mathrm{bulk}}L+\epsilon_{\alpha}+\epsilon_{\beta}+E_{\mathrm{NLIE}}\]
where the bulk energy constant $\mathcal{E}_{\mathrm{bulk}}$ and
the boundary energy terms $\epsilon_{\alpha,\beta}$ are known exactly
\cite{uv-ir}. The NLIE energy (\ref{eq:NLIE_energy}) is normalized
the same way as the boundary Lüscher formulae, so they can be compared
directly.

In the repulsive regime $\lambda<1$ one can calculate the leading
large volume asymptotics by substituting\[
Z\left(\theta\right)=2ML\sinh\theta+P\left(\theta\,|\, H_{0},H_{L}\right)\]
into (\ref{eq:NLIE_energy}) and taking $\eta\rightarrow\frac{\pi}{2}$.
The result is\begin{equation}
E=-M\int_{-\infty}^{\infty}\frac{d\theta}{4\pi}\cosh\theta\left(K_{\alpha}^{s\bar{s}}\left(-\theta\right)K_{\beta}^{\bar{s}s}\left(\theta\right)+K_{\alpha}^{\bar{s}s}\left(-\theta\right)K_{\beta}^{s\bar{s}}\left(\theta\right)\right)\mathrm{e}^{-2ML\cosh\theta}\label{eq:sol_antisol_asymp}\end{equation}
where\[
K_{\beta}^{s\bar{s}}\left(\theta\right)=R_{\alpha}\left(i\frac{\pi}{2}-\theta\right)_{s}^{s}\quad,\quad K_{\beta}^{\bar{s}s}\left(\theta\right)=R_{\beta}\left(i\frac{\pi}{2}-\theta\right)_{\bar{s}}^{\bar{s}}\]
in terms of the soliton and antisoliton reflection factors found by
Ghoshal and Zamolodchikov \cite{gz}. We made use of the relation
\cite{lmss,abr}\[
P\left(\theta\,|\, H_{0},H_{L}\right)=i\log S_{++}\left(2\theta\right)-i\log R_{\alpha}\left(\theta\right)_{s}^{s}-i\log R_{\beta}\left(\theta\right)_{s}^{s}+\pi\]
(where $S_{++}$ is the soliton-soliton phase shift) and also of bulk
crossing symmetry and boundary crossing-unitarity to bring the result
to the form (\ref{eq:sol_antisol_asymp}). This is in perfect agreement
with the result (\ref{eq:nondiag_two_part_corr}) obtained from the
cluster expansion. 

The above procedure does not work in the attractive regime because
the analyticity of $Z$ does not extend far enough to allow taking
the limit $\eta\rightarrow\frac{\pi}{2}$. In principle one should
be able to recover the cluster expansion analytically from the NLIE
expression (\ref{eq:NLIE_energy}) even in the attractive regime;
however, this requires more detailed understanding of the analytic
structure of the NLIE (\ref{eq:DNLIE}) than what we have at present,
and is out of the scope of the present paper. 

Instead we performed numerical comparison of NLIE results to the boundary
Lüscher formulae for several values of the coupling and the boundary
parameters. An example is shown in figure \ref{cap:ddv_comp} below,
where $\lambda=4$ and so both the first and the second breathers
are present in the bulk spectrum. Varying the boundary values of the
sine-Gordon field we investigated the cases when one expects the asymptotics
be given by the first or second breathers separately. The results
are summarized in figure \ref{cap:ddv_comp}, where $\log(|E|/M)$
is plotted against the dimensionless $l=ML$ with $M$ being the soliton
mass. (The sign of the exact ground state energy always matched the
sign of the appropriate asymptotic correction as calculated from (\ref{eq:m1_leading})
and (\ref{eq:m2_leading})). %
\begin{figure}
\begin{center}\includegraphics{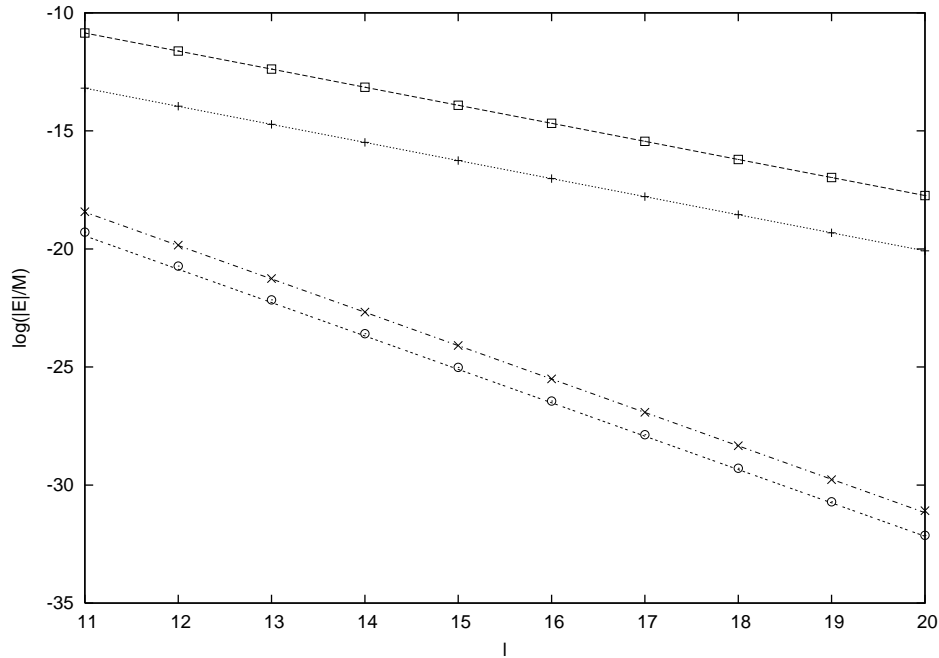}\end{center}

\caption{\label{cap:ddv_comp}Comparison of the Dirichlet NLIE with the boundary
one-particle contributions}
\end{figure}

For simplicity, we parameterize the boundary values $\Phi_{z}$, $z=0,\, L$
at the two ends of the strip in the following way\[
\Phi_{z}=\frac{2\pi}{\beta}\phi_{z}\]
and so $\eta_{z}=-(-1)^{z/L}\pi(1+\lambda)\phi_{z}$. The points marked
by empty boxes, $+$-s, $\times$-s and dotted circles are the exact
ground state values computed from the NLIE when the boundary values
of the sine-Gordon fields on the upper and lower ends of the strip
are $\phi_{0}=-0.17$ $\phi_{L}=0.1$, $\phi_{0}=0.017$ $\phi_{L}=0.1$,
$\phi_{0}=0$ $\phi_{L}=0$, and $\phi_{0}=-0.08$ $ \phi_{L}=0$,
respectively. The two upper continuous lines are the predictions of
the first breather's asymptotic contribution (\ref{eq:m1_leading})
(with $\vartheta\rightarrow\infty$ to describe the Dirichlet case),
while the two lower continuous lines are the predictions coming from
the second breather's asymptotic contribution (\ref{eq:m2_leading})
when at one end of the strip $\eta_{L}=0$: \[
E_{\alpha0}^{(2)}=-m_{2}\frac{g_{2}(\eta_{0},\infty)}{2}\frac{g_{2}\left(0,\infty\right)}{2}e^{-2m_{2}L}=-m_{2}\frac{\tan\left(\frac{\pi}{4\lambda}+\frac{\eta_{0}}{2\lambda}\right)\tan\left(\frac{\pi}{4\lambda}-\frac{\eta_{0}}{2\lambda}\right)}{\tan\left(\frac{\pi}{2\lambda}\right)\tan\left(\frac{\pi}{4}+\frac{\pi}{2\lambda}\right)}\, e^{-2m_{2}L}.\]
 The agreement with the NLIE data strongly supports the formulae derived
from the cluster expansion. Although for the graphical comparison
it is enough to plot only the second breather one-particle term, in
order to make the agreement with the NLIE numerically satisfactory
the two-particle contribution from the first breather also has to
be included. The detailed numerical comparison is performed in the
appendix and an excellent agreement is found. There's no such problem
with the first breather term, however, because it is possible to go
to large enough volume so that it is still numerically measurable
and at the same time dominates over the subleading corrections by
several orders of magnitude.

\section{Classical limit of scaling functions}

Motivated by the success of Mussardo et al. \cite{mrs_pbc} who showed
that in the case of periodic boundary conditions the (semi)classical
limit of L\"{u}scher's finite size correction to the sine-Gordon
soliton mass coincides with the infrared limit ($l=mL\rightarrow\infty$)
of the soliton's classical energy in finite volume, we show next that
the classical limit(s) of the previous L\"{u}scher type corrections
to the ground state energy in the boundary sine-Gordon theory on the
strip coincide with the asymptotic corrections to the classical energies
of the various static (ground state) solutions of the model.

\noindent In refs. \cite{bpt1,mrs} collectively the following static
solutions are described in details in terms of Jacobi elliptic functions:

\noindent $\bullet$ The ground state (+) and the solitonic excited
solution (-) with the $\Phi(0)=\Phi_{0}^{D}$, $\Phi(L)=0$ Dirichlet-Dirichlet
(DD) boundary conditions where in the (+) {[}(-){]} case the sine-Gordon
field $\Phi(x)$ decreases {[}increases{]} monotonically from $\Phi_{0}^{D}$
at $x=0$ to $0$ {[}$2\pi/\beta${]} at $x=L$; 

\noindent $\bullet$ the ground state (A) and the first excited state
(B) type solutions in the zero topological charge sector with the
\[
0<\varphi_{0}<\frac{\pi}{2},\qquad0\leq\varphi_{L}\leq\pi-\varphi_{0},\quad\varphi_{0,L}=\frac{\beta\Phi_{0,L}^{D}}{2},\]
 DD boundary conditions, where, for the case (A) {[}(B){]}, $\Phi(x)$
decreases {[}increases{]} from $\Phi_{0}^{D}$ to some turning point
$\epsilon$ {[}$\frac{2\pi}{\beta}-\tilde{\epsilon}${]} then increases
{[}decreases{]} from these values to $\Phi_{L}^{D}$; 

\noindent $\bullet$ the ground state (-1) of the $n=-1$ topological
charge sector with the \[
0<\varphi_{0}<\frac{\pi}{2},\qquad\frac{\beta\Phi(L)}{2}=\varphi_{L}-\pi\]
 DD boundary conditions, when $\Phi(x)$ decreases monotonically from
positive to negative values, 

\noindent $\bullet$ and finally the ground state with mixed boundary
conditions (DN): Dirichlet at $x=0$ and a one parametric ($M_{L}$)
perturbed Neumann type b.c. at $x=L$ \[
\frac{\beta}{2}\Phi(0)=\varphi_{0},\qquad\frac{\beta}{2}\Phi^{\prime}(x)|_{x=L}=-m\mathcal{A}^{-1}\sin(\varphi_{L}),\qquad\mathcal{A}^{-1}=\frac{M_{L}\beta^{2}}{4m}.\]

For all these static solutions the dimensionless width of the strip
$l=mL$ and the energies of the solutions can be expressed in terms
of the integrals \[
l(a,b,C)=\int\limits _{a}^{b}\frac{dv}{\sqrt{\sin^{2}v+C}},\qquad I(a,b,C)=\int\limits _{a}^{b}dv\sqrt{\sin^{2}v+C}.\]
 Indeed in the $\Phi_{L}^{D}=0$ Dirichlet-Dirichlet case for the
(+) solution one obtains \[
l=l(0,\varphi_{0},C),\qquad E_{+}(\varphi_{0},l)=\frac{4m}{\beta^{2}}[I(0,\varphi_{0},C)-Cl/2],\]
 while for the (-) solution \[
l=l(\varphi_{0},\pi,C),\qquad E_{-}(\varphi_{0},l)=\frac{4m}{\beta^{2}}[I(\varphi_{0},\pi,C)-Cl/2]\]
 results. In the case of general DD boundary conditions, in the $n=0$
sector, for the type (A) (ground state) solution \[
l=\sum\limits _{i=0,\,\, L}l(\epsilon,\varphi_{i},-\sin^{2}\epsilon),\qquad E_{A}(\varphi_{0},\varphi_{L},l)=\frac{4m}{\beta^{2}}[\sum\limits _{i=0,\,\, L}I(\epsilon,\varphi_{i},-\sin^{2}\epsilon)+\frac{l}{2}\sin^{2}\epsilon]\]
 is obtained, while for the type (B) (excited state) solution \[
l=\sum\limits _{i=0,\,\, L}l(\varphi_{i},\pi-\epsilon,-\sin^{2}\epsilon),\qquad E_{B}(\varphi_{0},\varphi_{L},l)=\frac{4m}{\beta^{2}}[\sum\limits _{i=0,\,\, L}I(\varphi_{i},\pi-\epsilon,-\sin^{2}\epsilon)+\frac{l}{2}\sin^{2}\epsilon]\]
 is found. In the $n=-1$ sector of the general DD boundaries for
the only ground state we obtain \[
l=l(0,\varphi_{0},C)+l(0,\pi-\varphi_{L},C),\qquad E^{(-1)}(\varphi_{0},\varphi_{L},l)=\frac{4m}{\beta^{2}}[I(0,\varphi_{0},C)+I(0,\pi-\varphi_{L},C)-Cl/2].\]
 Finally for the ground state of the DN problem \[
l=l(\varphi_{L},\varphi_{0},\tilde{C}),\qquad E^{(DN)}(\varphi_{0},\mathcal{A},l)=\frac{4m}{\beta^{2}}[-\tilde{C}\frac{l}{2}+I(\varphi_{L},\varphi_{0},\tilde{C})-\mathcal{A}^{-1}\cos(\varphi_{L})]\]
 is found with $\tilde{C}=-(1-\mathcal{A}^{-2})\sin^{2}\varphi_{L}$.
These expressions give the $l$ dependence of the energies in a parametric
form: in all cases the integration constant(s) $C$ or $\epsilon$
and $\varphi_{L}$ are -- at least in principle -- determined as functions
of $l$ (and the boundary parameters) from the first equations and
using them in the second equations gives the $l$ dependence of the
energy expressions.

The asymptotic form of the energies is obtained in the following way:
first in each case the integrals defining $l$ are expressed as an
appropriate linear combination of complete and incomplete elliptic
integrals. The $l\rightarrow\infty$ limit corresponds to the limit
when the parameter of these elliptic integrals goes to unity -- sometimes
this is also correlated with the amplitude of the incomplete one going
to $\pi/2$. Using the relevant asymptotic expansions of the elliptic
integrals we determine the asymptotic $C(l)$ ($\epsilon(l)$, $\varphi_{L}$).
In the second step the integrals appearing in the expressions for
the energy are determined by realizing that their derivatives with
respect to $C$ ($\epsilon$, $\varphi_{L}$) can be related to $l(a,b,C)$;
viewing this relation as a differential equation we integrate it (the
constant is fixed by the known value of $I(a,b,0)$). In this way
the following expressions are obtained at $l\rightarrow\infty$: \[
E_{+}\left(\varphi_{0},l\right)\sim\frac{4m}{ \beta^{2}}\left(1-\cos\varphi_{0}+4 \tan^{2}\frac{\varphi_{0}}{2}\,\mathrm{e}^{-2l}\right),\]
 \[
E_{-}\left(\varphi_{0},l\right)\sim\frac{4m}{ \beta^{2}}\left(1+\cos\varphi_{0}+4 \cot^{2}\frac{\varphi_{0}}{2}\,\mathrm{e}^{-2l}\right),\]
 \[
E_{A}\left(\varphi_{0},\varphi_{L},l\right)\sim\frac{4m}{\beta^{2}}\left(2-\cos\varphi_{0}-\cos\varphi_{L}-8\tan\frac{ \varphi_{0}}{2}\tan\frac{ \varphi_{L}}{2}\mathrm{e}^{-l}\right),\]
 \[
E_{B}\left(\varphi_{0},\varphi_{L},l\right)\sim\frac{4m}{\beta^{2}}\left(2+\cos\varphi_{0}+\cos\varphi_{L}-8\cot\frac{ \varphi_{0}}{2}\cot\frac{ \varphi_{L}}{2}\mathrm{e}^{-l}\right),\]
 \[
E^{(-1)}\left(\varphi_{0},\varphi_{L},l\right)\sim\frac{4m}{\beta^{2}}\left(2-\cos\varphi_{0}+\cos\varphi_{L}+8\tan\frac{ \varphi_{0}}{2}\tan\frac{ \pi-\varphi_{L}}{2}\mathrm{e}^{-l}\right),\]
 \begin{equation}
E^{(DN)}(\varphi_{0},\mathcal{A},l)\sim\frac{4m}{\beta^{2}}\left(1-\cos\varphi_{0}-{\mathcal{A}}^{-1}-4\tan^{2}\frac{\varphi_{0}}{2}\frac{1-\mathcal{A}^{-1}}{1+\mathcal{A}^{-1}}\mathrm{e}^{-2l}\right).\label{eq:semicl_energies}\end{equation}
The constant ($l$-independent) terms of these energy expressions
are the classical energies of the static solitons/antisolitons forming
the asymptotic ground state (for more details see \cite{bpt1,mrs}).
Note that some of the asymptotic corrections decrease with $\mathrm{e}^{-l}$,
while the others with $\mathrm{e}^{-2l}$. The fact that the asymptotic
correction is negative in the $n=0$ and positive in the $n=-1$ sectors
can be understood by realizing that in the former sector the asymptotic
solution is a superposition of a soliton and an antisoliton (whose
interaction is attractive) while in the latter it is a superposition
of two solitons (whose interaction is repulsive). The sign and magnitude
of these asymptotic corrections were also verified numerically by
comparing to the exact solution of the classical field equations.

We now proceed to show that these asymptotic corrections are indeed
the classical limits of the appropriate generalized Lüscher formulae. 

It is very important to realize that the classical limit (\ref{eq:eta_theta_scaling})
is not identical to taking $\beta\rightarrow0$, in which almost all
interesting quantities diverge (this limit cannot even be taken in
the classical formulae for the ground state energies). The general
form of the semiclassical expansion for the ground state energy can
be obtained from the Euclidean path integral as the limit\[
E(L)=\lim_{T\rightarrow\infty}-\frac{1}{\hbar T}\log\int D\Phi\exp\left\{ -\frac{1}{\hbar}\int_{-T/2}^{T/2}d\tau\, L_{E}(\Phi)\right\} \]
where the Euclidean Lagrangian is \begin{eqnarray*}
L_{E}\left(\Phi\right) & = & \int_{0}^{L}dx\left(\frac{1}{2}\left(\partial_{\tau}\Phi\right)^{2}+\frac{1}{2}\left(\partial_{x}\Phi\right)^{2}+\frac{m^{2}}{\beta^{2}}\cos\beta\Phi\right)\\
 & + & M_{L}\left[1-\cos\frac{\beta}{2}\left(\Phi(L,\tau)-\Phi_{L}\right)\right]+M_{0}\left[1-\cos\frac{\beta}{2}\left(\Phi(0,\tau)-\Phi_{0}\right)\right]\end{eqnarray*}
 Scaling out the coupling constant and introducing dimensionless variables\[
\Phi=\beta\bar{\Phi}\quad,\quad\bar{x}=mx\quad,\quad\bar{\tau}=m\tau\]
standard semiclassical expansion around the classical solution yields\begin{equation}
E(L)=\frac{m}{\beta^{2}}\left(F\left(mL,\frac{M_{0}\beta^{2}}{m},\frac{M_{L}\beta^{2}}{m},\beta\Phi_{0},\beta\Phi_{L}\right)+O\left(\hbar\beta^{2}\right)\right)\label{eq:semicl_expansion}\end{equation}
where $F$ is a dimensionless function of its arguments. The first
term is just the classical energy, the large volume asymptotics of
which is given in (\ref{eq:semicl_energies}) and is proportional
to $\beta^{-2}$. All the quantum corrections are regular as $\beta\rightarrow0$,
and expansion in powers of $\hbar$ corresponds to expansion in $\beta^{2}$.
As a result, the classical limit corresponds to isolating the leading
$\beta^{-2}$ term in the expansion around $\beta=0$. Thus it is
only meaningful to take the large volume asymptotic first and perform
the classical limit by considering only the leading terms for small
$\beta$.

The volume dependence of the finite size correction is determined
by the type of the particle giving the leading contribution: for terms
of the form $\mathrm{e}^{-l}$ it is the first breather, while in
the case of $\mathrm{e}^{-2l}$ the second (note that $m_{2}=2m_{1}=2m$
in the classical limit). 

The UV-IR relations (\ref{eq:UVIR_relation}) imply that the Dirichlet
boundary condition is obtained as the $\vartheta\:\rightarrow\:\infty$
limit of the general one, while the perturbed Neumann BC as the $\vartheta=0$
($\mathcal{A}^{-1}<1$) or $\eta=0$ ($\mathcal{A}^{-1}>1$) case.
Using the classical UV-IR relation (\ref{eq:classical_uvir}) one
finds that in the Dirichlet case \[
\eta_{\mathrm{cl},0}=-\varphi_{0}=-\frac{\beta\Phi_{0}}{2}\quad,\quad\eta_{\mathrm{cl},L}=\varphi_{L}=\frac{\beta\Phi_{L}}{2}\]
while for the perturbed Neumann BC\[
\cos\eta_{\mathrm{cl}}=\mathcal{A}^{-1}\quad(\mathcal{A}^{-1}<1)\qquad,\qquad\cosh\eta_{\mathrm{cl}}=\mathcal{A}^{-1}\quad(\mathcal{A}^{-1}>1)\]
results. 

Using the above identifications, the classical limit of the breather
one-particle term (\ref{eq:m1_leading}) becomes\[
-\frac{32m}{\beta^{2}}\tan\frac{ \varphi_{0}}{2}\tan\frac{ \varphi_{L}}{2}\mathrm{e}^{-l}\]
which indeed agrees with $E_{A}$.

To interpret $E_{B}$ that belongs to the first excited boundary state
let us recall that the reflection amplitudes on it can be obtained
by the substitution \begin{equation}
\eta\,\rightarrow\,\bar{\eta}=\pi\left(\lambda+1\right)-\eta\label{eq:etabar}\end{equation}
 from the ground state ones \cite{md,bptt}. Therefore making this
substitution in (\ref{eq:m1_leading}) gives the Lüscher type correction
to the first excited state energies. Since in the classical limit
this substitution simplifies to $\eta_{\mathrm{cl}}\,\rightarrow\,\pi-\eta_{\mathrm{cl}}$,
it is clear that $E_{B}$ does coincide with the classical limit of
(\ref{eq:m1_leading}). Similar reasoning can be used in the case
of $E^{(-1)}$.

For the corrections of order $\mathrm{e}^{-2l}$, however, things
are more complicated. Let us take the example of $E_{+}$, the other
cases can be treated in the same way. Taking the classical limit of
the second breather one-particle term (\ref{eq:m2_leading}) gives\begin{equation}
\frac{32m}{ \beta^{2}}\tan^{2}\frac{\varphi_{0}}{2}\,\mathrm{e}^{-2l}\label{eq:2br_semicl}\end{equation}
which is off by a factor of $2$. Fortunately this is not the end
of the story. As we already noted above, in the classical limit the
masses of the first and second breathers are given by \[
m_{1}=m\quad,\quad m_{2}=2m\]
and therefore the first breather two-particle term is exactly of same
order as the second breather one-particle term. Let us examine the
general formula for the two-particle contribution for a particle of
mass $m$ (\ref{eq:two_part_corr})\[
-m\int_{-\infty}^{\infty}\frac{d\theta}{4\pi}\cosh\theta\bar{K}_{\alpha}\left(\theta\right)K_{\beta}\left(\theta\right)\mathrm{e}^{-2mL\cosh\theta}=-m\mathrm{e}^{-2mL}\int_{- \infty}^{\infty}\frac{d\theta}{4\pi}f(\theta)\mathrm{e}^{-2mL\theta^{2}/2}\]
where we introduced \[
f(\theta)=\cosh\theta\bar{K}_{\alpha}\left(\theta\right)K_{\beta}\left(\theta\right)\mathrm{e}^{-2mL(\cosh\theta-1-\theta^{2}/2)}\]
Since $f(\theta)$ is an even function, it expands in even powers
of $\theta$ around $0$\[
f(\theta)=f_{0}+f_{1}\theta^{2}+\dots\]
and so the two-particle contribution can be written as (for $mL\rightarrow0$)\begin{equation}
-\frac{m}{4\sqrt{\pi mL}}\mathrm{e}^{-2mL}\left(f_{0}+\sum_{n=1}^{\infty}\frac{(2n-1)!!}{(2mL)^{n}}f_{n}\right)\label{eq:wrong_series}\end{equation}
At first sight this does not appear to be enough to rescue us, since
the two-particle correction is further suppressed by a factor of $L^{-1/2}$.
However, it is an interesting and nontrivial issue whether the asymptotic
expansion (\ref{eq:wrong_series}) can be trusted in the semiclassical
domain. 

The relevant integral in our case is \begin{equation}
-m_{1}\int_{-\infty}^{\infty}\frac{d\theta}{4\pi}\cosh\theta\log\left(1+\chi\left(\theta\right)\mathrm{e}^{-2m_{1}L\cosh\theta}\right)\sim-m_{1}\int_{-\infty}^{\infty}\frac{d\theta}{4\pi}\cosh\theta\chi\left(\theta\right)\mathrm{e}^{-2m_{1}L\cosh\theta}\label{eq:b1_2pszingint}\end{equation}
where \[
\chi\left(\theta\right)=R_{|\rangle}^{(1)}\left(i\frac{\pi}{2}-\theta\right)_{0}R_{|\rangle}^{(1)}\left(i\frac{\pi}{2}+\theta\right)_{L}\]
and for the case of $E_{+}$ the boundary parameters are $\eta_{0}=\varphi_{0}\left(\lambda+1\right)$,
$\eta_{L}=0$ and $\vartheta_{0}=\vartheta_{L}=\infty$.

Using the known reflection factors, the coefficients $f_{i}$ can
be explicitely calculated to have the following behaviour for small
$\beta$\[
f_{n}\sim\lambda^{2(n+1)}\sim\frac{1}{\beta^{4(n+1)}}\]
which means that in the classical limit the series expansion (\ref{eq:wrong_series})
is invalid.

The reason why the coefficients $f_{n}$ diverge and the integral
(\ref{eq:b1_2pszingint}) must be treated very carefully is that the
value of $\chi(\theta)$ and all of its derivatives at $\theta=0$
go to infinity in the classical limit. The dangerous contribution
comes from the particular block \[
\left(\frac{3}{2}+\frac{1}{2\lambda}\right)\]
in the reflection factor (\ref{b1refl}) and can be isolated from
$\chi$ as follows\[
\chi\left(\theta\right)=\frac{\cosh\theta+\cos\frac{\pi}{2\lambda}}{\cosh\theta-\cos\frac{\pi}{2\lambda}}\chi_{0}\left(\theta\right)\]
where $\chi_{0}$ is a function with a smooth classical limit at the
origin\[
\chi_{0}(0)=\left(\frac{\cos\frac{\pi}{4\lambda}-\sin\frac{\pi}{4\lambda}}{\cos\frac{\pi}{4\lambda}+\sin\frac{\pi}{4\lambda}}\tan\frac{\eta_{0}}{2\lambda}\right)^{2}\sim\tan^{2}\frac{\varphi_{0}}{2}+O\left(\beta^{2}\right)\]
while the function \[
g(\theta)=\frac{\cosh\theta+\cos\frac{\pi}{2\lambda}}{\cosh\theta-\cos\frac{\pi}{2\lambda}}\]
has the Taylor coefficients\[
\frac{g^{(2n)}(0)}{(2n)!}=(-1)^{n}4\left(\frac{2\lambda}{\pi}\right)^{2(n+1)}\left(1+O\left(\frac{1}{\lambda^{2}}\right)\right)\]
for large $\lambda$. As a result, the series (\ref{eq:wrong_series})
takes the following form for small $\beta$\[
-\frac{256m}{\sqrt{\pi mL}\beta^{4}}\tan^{2}\frac{\varphi_{0}}{2}\mathrm{e}^{-2mL}\left(1+\sum_{n=1}^{\infty}(2n-1)!!\left(-\frac{128}{mL\beta^{4}}\right)^{n}\right)\]
which is a power series with coefficients that grow factorially. It
is also obvious that it does not have the right $\beta$ dependence
displayed in eqn. (\ref{eq:semicl_expansion}), and the higher terms
show more and more singular behaviour at $\beta=0$.

In order to get the correct asymptotics, we now evaluate the leading
behaviour of the integral (\ref{eq:b1_2pszingint}) in the semiclassical
regime carefully. First we substitute $x=\sinh\theta$ for the integration
variable. For large $L$, the integrand is concentrated around the
origin and therefore we can evaluate the dominant contribution by
keeping only the leading terms around $x=0$:\[
-m_{1}\int_{-\infty}^{\infty}\frac{dx}{4\pi}\log\left(1+\left(\frac{1+\cos\frac{\pi}{2\lambda}+x^{2}/2}{1-\cos\frac{\pi}{2\lambda}+x^{2}/2}-1\right)\chi_{0}(0)\mathrm{e}^{-2m_{1}L}\right)\]
where a subtraction of $-1$ in the coefficient of $\chi_{0}(0)\mathrm{e}^{-2m_{1}L}$
was applied to make the integral convergent at the plus and minus
infinities (this only affects subleading terms -- of order $\beta^{0}$
-- as the singular behaviour around $x\sim0$ is unchanged). Now the
integral can be evaluated using (\ref{eq:integral_formula})\[
-m_{1}\left[\sqrt{\sin^{2}\frac{\pi}{4\lambda}+\cos\frac{\pi}{2\lambda}\chi_{0}(0)\mathrm{e}^{-2m_{1}L}}-\sin\frac{\pi}{4\lambda}\right]\sim-\frac{m_{1}}{2}\chi_{0}(0)\frac{\cos\frac{\pi}{2\lambda}}{\sin\frac{\pi}{4\lambda}}\mathrm{e}^{-2m_{1}L}\]
using that $L$ is very large. 

Taking the classical limit of small $\beta$ (i.e. large $\lambda$)
and keeping only the leading term in $\beta$ we obtain\[
-\frac{16m}{ \beta^{2}}\tan^{2}\frac{\varphi_{0}}{2}\,\mathrm{e}^{-2mL}\]
which must be added to the second breather one-particle contribution
(\ref{eq:2br_semicl}), thereby giving perfect agreement with the
classical formula for $E_{+}$. The same argument works for $E^{(DN)}$,
while $E_{-}$ can be obtained from $E_{+}$ by the substitution (\ref{eq:etabar}).

\section{Conclusions}

In this paper we succeeded in deriving the boundary analogue of Lüscher's
analytic formulae for finite size corrections (which were originally
obtained for periodic boundary conditions). The consistency of these
boundary Lüscher formulae was checked against boundary TBA, NLIE and
classical field theoretic results. The matching of all these different
approaches provided ample evidence for the correctness of the formulae,
and also a very strong argument for the universality of the relation
\begin{equation}
\tilde{g}_{\alpha}=\frac{g_{\alpha}}{2}\label{eq:gg_relation}\end{equation}
between the one-particle coupling $g_{\alpha}$ in the reflection
factors and the one-particle amplitude $\tilde{g}_{\alpha}$ in the
corresponding boundary state, which was first proposed for the boundary
Lee-Yang model in \cite{dptw_onepoint} on the basis of matching the
form factor expansion of one-point function against truncated conformal
space data. Our numerical data based on the Dirichlet NLIE and the
matching with the classical finite size corrections show that this
relation is valid in boundary sine-Gordon theory for very general
values of parameters. The argument based on the boundary TBA shows
that it is also valid for any theory with a diagonal bulk and boundary
scattering theory. Based on these results we conjecture that this
relation is indeed universal.

Using the relation (\ref{eq:gg_relation}), the boundary Lüscher formulae
proposed in this paper make it possible to compute the leading finite
size corrections based only on data from the scattering amplitudes,
even in non-integrable theories. It is interesting to note a deep
analogy between the boundary Lüscher formulae (\ref{eq:two_part_corr},\ref{eq:luscher_onepart})
and the particle mass correction (\ref{eq:original_luscher}) for
the periodic case. In both cases the generic correction can be written
in terms of an integral over the relevant scattering amplitude, and
the two integrals share many other similarities in structure. Besides
that, the possible pole terms are also very similar. Therefore we
expect that a quantum field theoretic derivation of our results along
the lines of \cite{luscher} must exist. Such a derivation would be
very interesting because it would permit generalization to any space-time
dimensions (similarly to the case of periodic boundary condition \cite{luscher}),
while the cluster expansion technique is restricted at the moment
to $1+1$ dimensional field theories, albeit not necessarily integrable
ones.

While the exponential dependence of the finite size corrections on
the volume and the particle masses is obvious from general principles
of field theory and TBA/NLIE, there are some highly nontrivial aspects
of the results obtained here which we would like to stress. First
of all, the relation (\ref{eq:gg_relation}) which was originally
a (numerical) observation made in a particular model (Lee-Yang) in
\cite{dptw_onepoint}, is shown to be universal by several pieces
of evidence (analytic and numeric). Second, we now have a universal
analytic form of the exponential finite size corrections expressed
in terms of the infinite volume scattering theory, and also a regularized
form for the two-particle contribution (eqn. (\ref{eq:correct_as}))
when the naive expression (\ref{eq:two_part_corr}) is divergent.
Third, we have done extensive consistency checks of the results, and
in particular we have shown that they agree with classical field theory,
but the classical limit of finite size corrections must be treated
very carefully.

In more practical terms, the boundary Lüscher formulae are very useful
for extracting boundary scattering data from finite size effects obtained
by any numerical or analytic method like lattice field theory, truncated
conformal space or Bethe Ansatz based approaches (TBA, NLIE). Alternatively,
when the scattering theory is explicitely known they can be applied
to test the validity of numerical methods or conjectured TBA or NLIE
equations. Such applications are presently being actively pursued
by the authors.

\subsection*{Acknowledgments.}

The authors would like to thank G. Watts and C. Rim for illuminating
discussions, and R. Nepomechie for pointing out a mistaken (and important)
sign. GT would like to thank the Department of Mathematics, King's
College London, for their hospitality during the course of the work.
This research was partially supported by the EC network {}``EUCLID'',
contract number HPRN-CT-2002-00325, and Hungarian research funds OTKA
D42209, T037674, T034299 and T043582. GT was also supported by a Széchenyi
István Fellowship.

\appendix
\renewcommand{\theequation}{\hbox{\normalsize\Alph{section}.\arabic{equation}}}

\renewcommand{\thefigure}{\hbox{\normalsize\Alph{section}.\arabic{figure}}}

\renewcommand{\thetable}{\hbox{\normalsize\Alph{section}.\arabic{table}}}

\section{Numerical comparison to the NLIE}

We consider the first breather one-particle contribution first. We
normalize the energies and distances in units of the soliton mass
$M$:\begin{eqnarray}
\frac{E_{\alpha\beta}\left(L\right)}{M} & = & 2\sin\frac{\pi}{2\lambda}\frac{1+\cos\frac{\pi}{2\lambda}-\sin\frac{\pi}{2\lambda}}{1-\cos\frac{\pi}{2\lambda}+\sin\frac{\pi}{2\lambda}}\tan\frac{\eta_{0}}{2\lambda}\tan\frac{\eta_{L}}{2\lambda}\mathrm{e}^{-2\sin\frac{\pi}{2\lambda}l}+\dots\label{eq:dimless_b1_corr}\\
l & = & ML\nonumber \end{eqnarray}
and the boundary values of the field as \[
\Phi_{z}=\frac{2\pi}{\beta}\phi_{z}\quad,\quad\eta_{z}=-(-1)^{z/L}\phi_{z}\pi\left(1+\lambda\right)\]
Numerical data are obtained by fitting the NLIE results to a curve
of the form\[
\gamma\mathrm{e}^{-\alpha l}\]
Illustrative examples are given in table \ref{cap:b1_nlie}; the agreement
was tested for many other values of the parameters involved, with
results similar to those in the table.

\begin{table}
\begin{center}\begin{tabular}{|c|c|c|c|c|c|}
\hline 
&
$\phi_{0}$&
$\phi_{L}$&
$\alpha$&
$\gamma_{\mathrm{measured}}$&
$\gamma_{\mathrm{exact}}$\tabularnewline
\hline 
$\lambda=7/2$&
&
&
exact: $0.867767$&
&
\tabularnewline
\hline 
&
$0.05$&
$0.05$&
$0.867774$&
$-0.024529$&
$-0.024526$\tabularnewline
\hline 
&
$0.03$&
$0.07$&
$0.867812$&
$-0.020640$&
$-0.020625$\tabularnewline
\hline 
&
$0.07$&
$0.07$&
$0.867786$&
$-0.048403$&
$-0.048389$\tabularnewline
\hline
\hline 
$\lambda=5/2$&
&
&
exact: $1.175570$&
&
\tabularnewline
\hline 
&
$0.1$&
$0.2$&
$1.176765$&
$-0.197730$&
$-0.193904$\tabularnewline
\hline 
&
$0.05$&
$0.15$&
$1.175572$&
$-0.069682$&
$-0.069681$\tabularnewline
\hline 
&
$-0.05$&
$0.15$&
$1.175571$&
$0.069681$&
$0.069681$\tabularnewline
\hline
\end{tabular}\end{center}

\caption{\label{cap:b1_nlie} Comparing the first breather one-particle term
to NLIE. For each value of $\lambda$, we quote the exact value of
$\alpha=m_{1}/M=2\sin\frac{\pi}{2\lambda}$, which is compared to
the exponent extracted from the NLIE. $\gamma_{\mathrm{measured}}$
is the coefficient extracted from NLIE, while $\gamma_{\mathrm{exact}}$
is the one predicted by (\ref{eq:dimless_b1_corr}).}
\end{table}

The second breather one-particle term proves more tricky. To eliminate
the otherwise dominant first breather one-particle term we choose
$\phi_{L}=0$. However, the two-particle term for the first breather
gives a non-negligible contribution for all values of the parameters
that are accessible by iterating the NLIE. Therefore we can only compare
the NLIE results to the sum of the two terms. The first breather two-particle
term can be safely calculated from (\ref{eq:two_part_corr}) since
the integrand is regular. Table \ref{cap:b2_nlie} summarizes some
of the results obtained this way. The agreement between the NLIE and
the boundary Lüscher formula becomes worse with decreasing volume,
which is easy to understand since higher order terms of the cluster
expansion start playing an increasingly important role. Once again,
we have performed a rather extensive numerical check, of which the
table contains only a small sample.

\begin{table}
\begin{tabular}{|c|c|c|c|c|c|}
\hline 
$l$&
&
$10$&
$12$&
$14$&
$16$\tabularnewline
\hline 
$\lambda=7/2,\phi_{0}=0.1$&
2 $B_{1}$&
-4.20321e-10&
-1.17697e-11&
-3.35392e-13&
-9.67996e-15\tabularnewline
\hline 
&
$B_{2}$&
-1.87457e-09&
-8.21717e-11&
-3.60199e-12&
-1.57893e-13\tabularnewline
\hline 
&
$\sum$&
-2.29490e-09&
-9.39414e-11&
-3.93738e-12&
-1.67572e-13\tabularnewline
\hline 
&
NLIE&
-2.29710e-09&
-9.39742e-11&
-3.93787e-12&
-1.67577e-13\tabularnewline
\hline
\hline 
$\lambda=9/2,\phi_{0}=0.013$&
2 $B_{1}$&
-6.33781e-09&
-3.37019e-10&
-1.84416e-11&
-1.03073e-12\tabularnewline
\hline 
&
$B_{2}$&
-1.31031e-07&
-1.00170e-08&
-7.65776e-10&
-5.85417e-11\tabularnewline
\hline 
&
$\sum$&
-1.37369e-07&
-1.03540e-08&
-7.84218e-10&
-5.95724e-11\tabularnewline
\hline 
&
NLIE&
-1.37425e-07&
-1.03550e-08&
-7.84232e-10&
-5.95718e-11\tabularnewline
\hline
\hline 
$\lambda=9/2,\phi_{0}=0.007$&
2 $B_{1}$&
-1.53201e-08&
-8.84074e-10&
-5.20261e-11&
-3.10504e-12\tabularnewline
\hline
&
$B_{2}$&
-5.51243e-08&
-4.21411e-09&
-3.22159e-10&
-2.46282e-11\tabularnewline
\hline
&
$\sum$&
-7.04444e-08&
-5.09819e-09&
-3.74185e-10&
-2.77333e-11\tabularnewline
\hline
&
NLIE&
-7.04671e-08&
-5.09861e-09&
-3.74193e-10&
-2.77334e-11\tabularnewline
\hline
\end{tabular}

\caption{\label{cap:b2_nlie} NLIE versus second breather one-particle term
combined with the first breather two-particle term}
\end{table}

\end{document}